\documentclass[12pt]{article}
\usepackage{amssymb,amsmath,amsthm}
\usepackage{rotating}
\usepackage{graphicx,comment}
\usepackage{color}

\excludecomment{omitext}

\begin{document}

% 12 FEBR 19 - pentru TIBA ??

\begin{center} 

{ \bf A strong contradiction in the  multi-layer Hele-Shaw model. } 

 Gelu Pa\c{s}a

\end{center}

{ \small
{\bf Abstract}. 
The  Saffman-Taylor  instability  occurs when a less viscous fluid is displacing 
a more viscous one in a rectangular Hele-Shaw cell. A surface tension on the interface
between the  two fluids is improving the  stability.
The multi-layer Hele - Shaw model, consisting  of $N$ intermediate  fluids with constant  
viscosities,  was studied   in some previous papers and very low growth constants
 were obtained for large  $N$.
 We prove that this  model leads us  to a significant instability, even if $N$  
is  very large. The maximum value of growth constants 
can  not  decrease under a certain value, not depending on the surface tensions on the interfaces. 
This contradiction with the Saffman-Taylor result  makes us have some doubts 
concerning the correctness of multi-layer model. 

%  28 feb 2019 LAPTOP 

\vspace{0.5cm}



\begin{center} 

{\bf   1. Introduction} 

\end{center}

We consider  a  Stokes flow  in a Hele-Shaw cell:
the narrow space between two parallel plates, first studied in 
 \cite{HS}.   The  velocities averages are verifying                                        
the   Darcy's law for the flow in  a porous 
medium  with the  permeability  $(b^2/12) $, where 
$b$ is the distance between the plates 
-  see  \cite{BE}, ~\cite{LAMB}

%%%%%%%%%%%%%%%%%%%%%%%%%%%%%%%%%%%%%%%%%

Saffman and Taylor   ~\cite{SAFF-TAY} proven the well-known result:
the interface between two imiscible fluids is unstable  when the 
displacing fluid  is less viscous. Moreover, the   growth 
rate   is unbounded with respect to   the wave numbers if the surface 
tension on the interface is missing.
A surface tension on the interface is limiting the range of 
 unstable   disturbances   - see the formula (11)  in \cite{SAFF-TAY}.
The fingering phenomenon (which appears for  unstable displacements) 
and the selection problem  in Hele-Shaw   displacements
 are  studied in a large number  of papers - see  ~\cite{HOM}, ~\cite{SAFF},
 ~\cite{XU} and references therein.  
In ~\cite{TA1} was studied the case  when the surface tension is very small.
Some singular effects due to the  zero-surface-tension problem  are 
studied in ~\cite{TA2}.

The optimization of  displacements in porous media were studied  in 
~\cite{AL-HUS-1},   ~\cite{AL-HUS-2},  ~\cite{CHEN},  ~\cite{DIAZ},  
~\cite{SUDARYANTO}. An intermediate fluid  with   variable  viscosity  
in a middle layer between the  displacing  fluids can minimize  the  
Saffman-Taylor instability  if surface tensions exist -  see 
experimental and numerical results  in \cite{GILJE}, ~\cite{GOR-HOM-1},  
~\cite{GOR-HOM-2},  ~\cite{SHAH},  ~\cite{SLOBOD},    ~\cite{UZOIGWE}.  
%%%%%%%%%%%%%%%%%%%%%%%%%%%%%%%%%%%%%%%%%%%%%%%%%%%%%%%%%%%%%%%%%%%%%%%%%
In   ~\cite{DAR-PA-1}, ~\cite{DAR-PA-2},  ~\cite{DAR-PA-3} are given 
theoretical results concerning  the linear stability  of such  three -
layer Hele-Shaw flow. Some  exact formulas of the growth constants  were 
given,  for variable  and  constant  intermediate viscosities.

%%%%%%%%%%%%%%%%%%%%%%%%%%%%%%%%%%%%%%%%%%%%%%%%%%%%%%%%%%%%%%%%%%%%%%%%%%%
The Hele-Shaw displacement  with  $N$  intermediate layers (the multi-layer 
Hele - Shaw model)  was studied  in ~\cite{DAR-2}, \cite {D6},  ~\cite{DAR-DING},  
\cite {D7}.  Only upper bounds  of the growth rates were obtained. 
In the case of  intermediate viscosities with  positive jumps in the flow 
direction, in ~\cite{DAR-2}  was proved that the corresponding growth rates  
tend to zero when the  number  of the intermediate  layers  is   
very large   and the  surface tensions  satisfy some   conditions.
%%%%%%%%%%%%%%%%%%%%%%%%%%%%%%%%%%%%%%%%%%%%%%%%%%%%%%%%%%%%%%%%%%%%%%%%%

In this paper we study   the multi-layer Hele - Shaw model with constant  
$N$ intermediate viscosities.  
A new  upper estimate  of the growth rates is obtained, for a bounded range 
of the wavenumbers and large $N$.    
We show that the maximum value of the growth constants {\it is  not depending 
on the surface tensions on the interfaces}  and  can not be less than  the 
difference between the viscosities of  the  initial displacing  fluids. 
Therefore the stability can not be improved, even if the surface tensions on 
the  interfaces are  very large.
This contradiction with the Saffman-Taylor result  makes us have some doubts 
concerning the correctness of multi-layer model.

 An important new element is given by  the  terms of Dirac type,  
appearing in the  estimates   of  the growth rates. These terms are related with  
the derivatives of the viscosity  across  the interfaces.

The paper is laid  out  as follows. 
In section 2   we describe the three-layer Hele-Shaw model with variable 
intermediate  viscosity, first studied in \cite{GOR-HOM-1}. 
In section 3   we get the formula  of the growth rates corresponding  to a
fourth-layer Hele-Shaw flow  with constant intermediate viscosities. 
This result is used in section 4,  for a model with  $N$ intermediate layers 
with constant  viscosities. We conclude in section 5.

\begin{center}

{\bf  2.  The three-layer Hele-Shaw model  with variable intermediate viscosity}.

\end{center}

The three-layer Hele-Shaw flow with variable  intermediate viscosity was first described in 
~\cite{GOR-HOM-1},  ~\cite{GOR-HOM-2}. The  cell is parallel with the $xOy$ plane. 
An intermediate region between the 
initial immiscible fluids is considered, where a  given amount of polymer-solute exists. The 
adsorption, dispersion and diffusion of the solute in the equivalent  porous medium are neglected.  
The intermediate viscosity can  be considered  as a  powers series  with respect 
to the concentration $C$ of the  polymer-solute   - see   ~\cite{FLORY},  ~\cite{GILJE}. 
For  a dilute  solute, the viscosity  is  a linear expression with respect to  $C$, then is
 invertible. We consider $\mu=12 \nu / b^2 $, where $\nu$ is the viscosity on the intermediate
region.  
The continuity equation for the solute   gives us the  ``continuity'' equation  
$\mu _t + u \mu_x  + v \mu_y = 0$,  where $(u,v)$ are the velocities and the indices $t, x,y$  
denote  the partial derivatives with respect to time and spatial variables. 

%%%%%%%%%%%%%%%%%%%%%%%%%%%%%%%%%%%
During the displacement process, the initial sharp interfaces change over time  and the finger
phenomenon appears. We consider small enough time intervals, to avoid large deformations of the 
initial interfaces.

 Mungan ~\cite{MUN} used an intermediate  polymer-solute with an exponentially- decreasing viscosity 
(from the front  interface) and obtained an  almost stable flow. 
 The displacements with variable   viscosity in Hele-Shaw cells   and   porous media  
are  studied in ~\cite{LOGGIA},  ~\cite{TALON}. 
On the page 3 of ~\cite{GEOLOGIC}  is considered a linear viscosity profile in a porous medium.

In this paper, the displacing and displaced fluids  are  denoted with the lower indices $_W$,   $_O$.  

Suppose the intermediate region  is the interval $ x \in (Ut - Q,  Ut) $, 
moving with the  constant   velocity $U$ far upstream.   We have three incompressible   fluids with 
viscosities  $\mu_W$ (displacing fluid),  $ \mu $ (intermediate  layer) and  $ \mu_O$ (displaced 
fluid). 
 In a large number of papers, the flow equations, quite similar with Darcy's 
law for flow in a porous medium,  are written in simpler form 
\begin{equation}\label{ZT004-0}
  p_{x} = - \mu_d u; \,\,  p_y = -\mu_d v;  
 \,\, p_z=0; \quad  u_{x} + v_y = 0;                                                          
\end{equation}                                                     
\begin{equation}\label{ZT004}
 \mu_d = \mu_W, \,\, x < Ut-Q; \,\,\,
 \mu_d = \mu, \,\,   x \in ( Ut - Q,   Ut) ; \,\,\,
 \mu_d = \mu_O, \,\, x > Ut. 
\end{equation}               
The  viscosities $\nu_W, \nu, \nu_O$   are given by
\begin{equation}\label{DIM-VISCO}
\mu_W = 12\nu_W /b^2, \,  \mu = 12 \nu /b^2,      \,  
\mu_O =  12 \nu_O/b^2,                                                                   
\end{equation}
and  the permeability of the equivalent porous medium is $b^2/12$. 
The velocities appearing in  \eqref{ZT004}  are the average of 
the real (effective)  fluid velocities - see  \cite{BE},  \cite {HS}, 
~\cite{LAMB}.

\vspace{0.25cm}	

 The  basic velocity and interfaces are  $ u=U, \,\, v=0; \quad  
\quad  x = Ut-Q, \,\, x =Ut $.

On the interfaces we consider the  Laplace's law: the 
pressure jump is given by the surface tension multiplied with 
the curvature of the  interface . The  component $u$ 
of the velocity is continuous  and  the interface is a 
material one. 

The basic interfaces are straight lines,  then 
the basic pressure $P$ is  continuous (but his gradient is not) 
and          
 \begin{equation}\label{BASIC-PRESS}
P_{x}= -  \mu_d U, \quad P_y=0.
\end{equation}

We use the     ``continuity'' equation for $\mu$ (see the 
end  of the first paragraph of this section), then the basic (unknown) 
 $\mu$ in the middle layer verifies  the equation
\begin{equation}\label{ZT004A}
 \mu_t + U \mu_{x} = 0.
\end{equation}
We introduce the moving reference frame
$ {\overline x}=x - Ut, \quad    \tau = t$.                
The equation  \eqref{ZT004A}  leads to $\mu_{\tau}=0$, then  
$\mu = \mu ({\overline x}) $. The middle region in the  moving  
reference frame is the segment  $ -Q <{\overline x} < 0 $. However,  
we  still use  the notation  $x, \,\, t$ instead of 
${\overline x}, \tau$.

\vspace{0.25cm}

The perturbations  $u', v', p', \mu' $  of the basic velocity, pressure 
and viscosity are  governed by the system  (see ~\cite{GOR-HOM-1})
\begin{equation}\label{ZT0040}
 p'_x = -\mu u' - \mu' U, \quad p'_y = -  \mu v',                
\end{equation} 
\begin{equation}\label{ZT005A}
 u'_x + v'_y = 0,
\end{equation}
\begin{equation}\label{ZT006A}
 \mu'_t + u' \mu_x = 0. 
\end{equation}
A Fourier decomposition for the perturbation $u'$  is used:
\begin{equation}\label{FOURIER-U}
 u'(x,y,t) = 
f(x)  [   \cos(ky) + \sin (ky)] e^{\sigma t}, \,\, k \geq 0,  
\end{equation}
where   $f(x), \sigma, k $ are the amplitude, the growth constant  and  
 the wave numbers.

\vspace{0.25cm}

The  velocity along the axis $Ox$ is continuous, then the amplitude 
$f(x)$ is  continuous. From \eqref{ZT0040} -  \eqref{FOURIER-U}  we get  
the Fourier decompositions  for the perturbations $v', p', \mu'$:
$$ v' = ( 1/k) f_x 
[-  \sin(ky) + \cos(ky)] e^{\sigma t},                                      $$
$$ p' = (\mu / k^2) f_x 
[-  \cos(ky) - \sin(ky)] e^{\sigma t},                                      $$                               
\begin{equation}\label{ZT007}
 \mu' = (-1/ \sigma) \mu_x f
[  \cos(ky) +  \sin (ky)]e^{\sigma t}.           
\end{equation} 
The cross derivation of the relations $\eqref{ZT0040}_1, 
\eqref{ZT0040}_2$    leads us to
\begin{equation}\label{Z-CROSS}
 \mu u'_y + \mu'_y U =  \mu_x v' + \mu v'_{x} .              
\end{equation}
From  \eqref{FOURIER-U}, $\eqref{ZT007}_1$,  \eqref{Z-CROSS}  
we get the  equation which  governs   the amplitude  $f$:
\begin{equation}\label{ZT008}
 -(\mu f_x)_x +  k^2 \mu f = \frac{1}{\sigma} U k^2 f \mu_x,  
 \quad \forall x \notin \{-Q,0 \}.
\end{equation}
Outside the intermediate region we have  constant viscosities, then   
\eqref{ZT008}  becomes 
$$ -f_{xx} + k^2 f = 0, \quad x \notin (-Q,0)                   $$
and in   the far field  we have
\begin{equation}\label{FAR-FIELD}
f(x) = \left \{  \begin{array}{c}
f(-Q) e^{ k(x+Q) }, \,\, \forall x \leq -Q; \\
f(0) e^{ -kx }, \,\, \forall x   \geq 0. 
 \end{array} \right.
\end{equation}
 
\vspace{0.25cm}

Suppose that a viscosity jump exists at a point $a$.
The  perturbed interface near $a$  is denoted by $\eta(a,y,t)$.
In the first approximation we have  $\eta_t = u$, therefore   
\begin{equation}\label{INTER001}
\eta(a,y,t) =  
(1/ \sigma)  f(a) 
[  \cos(k y) +  \sin(ky) ] e^{\sigma t }.
\end{equation}  
 The right and left limit values  of the pressure in 
the point   $a$ are  denoted  by $p^+(a), \quad p^-(a)$. 
 We use  $P$   in the point  $a$, the  Taylor first order  expansion  
of $P$ near $a$ and   $p'(a)$ given by   $\eqref{ZT007}_2$.   From 
\eqref{BASIC-PRESS}  it  follows  
$P_x^{+}(a) = -\mu^{+}(a)U $   and $P_x^{-}(a) = -\mu^{-}(a)U $, then  
we get 
$$  p^+(a) = P^+(a) + P^+_x(a) \eta + p'^+(a) =             $$  
\begin{equation}\label{INTER002}
 P^+(a)       
- \mu^+(a)  \{ \frac{U f(a)}{\sigma} +\frac{ f_x^+(a)}{k^2} \}
[ \cos(k y) +  \sin(ky) ]e^{\sigma t},      
\end{equation}              
\begin{equation}\label{INTER003}
p^-(a)  =   P^-(a)      
- \mu^-(a)  \{ \frac{U f(a)}{\sigma} +\frac{ f_x^-(a)}{k^2} \}
[ \cos(k y) +  \sin(ky) ]e^{\sigma t},       
\end{equation}               
The  Laplace's law is   $p^+(a) - p^-(a) = T(a) \eta_{yy}$,
where $T(a)$ is the surface tension
and  $\eta_{yy}$  is the approximate value  of the  curvature 
of  the perturbed  interface. As $P^-(a)=P^+(a)$ (see the line  before
\eqref{BASIC-PRESS}),  from the jump relation 
equations \eqref{INTER002} -\eqref{INTER003}   we get 
\begin{equation}\label{LAPLACE002}
- \mu^+(a)[\frac{Uf(a)}{\sigma}+  \frac{f_x^+(a)}{k^2}] +     
  \mu^-(a)[\frac{Uf(a)}{\sigma}+\frac{f_x^-(a)}{k^2}] =        
-  \frac{T(a)}{\sigma}f(a)k^2.
\end{equation}

\vspace{0.25cm}

 The growth constant  for three-layer case is obtained  as follows. 
We  multiply with $f$ in the amplitude equation  \eqref{ZT008}, we 
integrate on $(-Q,0)$ and obtain  
$$  - \int_{-Q}^0 (\mu f_x f)_x +  
\int_{-Q}^0 \mu f_x^2 +     k^2 \int_{-Q}^0 \mu f^2 =      
   \frac{k^2U}{\sigma} \int_{-Q}^0 \mu_x f^2.              $$
 We have not jumps of $\mu$ inside the intermediate region, then we 
get
$$
 \mu^+(-Q)f_x^+(-Q)f(-Q) -  \mu^-(0)f_x^-(0)f(0) +         $$
\begin{equation}\label{PRE-SIGMA}
 \int_{-Q}^0 \mu f_x^2 +     k^2 \int_{-Q}^0 \mu f^2 =   
\frac{k^2U}{\sigma} \int_{-Q}^0 \mu_x f^2.                   
\end{equation}
From the relations \eqref{FAR-FIELD}  we have   
\begin{equation}\label{PRE-SIGMA-B}    
f_x^-(-Q) =  k f_1:= kf(-Q), \quad  f_x^+(0)=-k f_{0}:=k f(0).           
\end{equation} 
Recall  $ \mu^-(-Q)= \mu_W, \,\,
\mu^+(0)= \mu_O$, then  from  \eqref{LAPLACE002}, \eqref{PRE-SIGMA},  
\eqref{PRE-SIGMA-B} it  follows
$$
 \sigma = \frac{ S_0 f_0^2 + S_1 f_1^2 + 
k^2 U \int_{-Q}^0 \mu_x f^2}
{\mu_Ok f^2_0 +  \mu_W kf^2_1 + I},                           $$
\begin{equation}\label{SIGMA001}
S_0 = k^2U[\mu]_0 - k^4 T_0,  \quad S_1 = k^2U[\mu]_1 - k^4 T_1,    
\quad    I =  \int_{-Q}^0 [ \mu f_x^2 + k^2 \mu f^2 ],           
\end{equation}
where $T_0, T_1$ are the surface tensions in $x=0, x=-Q$   and  
$$[\mu]_0=  (\mu^+ - \mu^-)_0= \mu_O-\mu^-(0),                $$
\begin{equation}\label{SIGMA001A}
  [\mu]_1=  (\mu^+ - \mu^-)_1= \mu^+(-Q) -\mu_W.                    
\end{equation}

\vspace{0.25cm}

{\it Remark 1}. 
From  \eqref{LAPLACE002} we can recover the  Saffman - Taylor  
formula
\begin{equation}\label{SIGMA_ST01} 
\sigma_{ST}=\frac{k U(\mu_O - \mu_W) -T(a) k^3}{\mu_O + \mu_W}. 
\end{equation}

Indeed,  we have 
$$\mu^+(a)=\mu_O; \quad \mu^-(a)=\mu_W;                          $$  
$$ f(x) = f(a) e^{ k(x-a) },  \quad  x \leq a \Rightarrow  \quad  
f_x^-(a)  =kf(a) ;                                              $$
$$ f(x) = f(a) e^{ -k(x-a) }, \quad  x \geq a \Rightarrow \quad                        
f_x^+(a)  =- kf(a).   \eqno \square                             $$

{\it Remark 2}.
It is possible  to inject   polymer-solutes  with constant 
concentrations  $c_1,c_2,..., c_N$  during some the time intervals    
$t_1, t_2, ..., t_N$.
We obtain a steady flow of $N$ thin layers of immiscible fluids  with 
{\it constant} viscosities  $\nu_i, \quad i=1,2,...,N$. This is the 
multi-layer model studied in  ~\cite{DAR-2},   ~\cite{DAR-DING},  
\cite {D7} , \cite {D6}.
$ \hfill \square $

\newpage

\begin{center}

{\bf  3.  The fourth-layer Hele-Shaw model  
 with constant intermediate  viscosities.}

\end{center}

Consider  two intermediate layers $(-Q, -Q/2)$ and $(-Q/2, 0)$ with 
constant $\mu$:
$$ \mu(x)=\mu_W, \,\, x< -Q, \quad \mu(x) =\mu_O, \,\, x>0,       $$ 
$$ \mu(x) = \mu_2,   \,\,\,  x \in   (-Q, -Q/2),  \quad  
   \mu(x) = \mu_1,  \,\,\,  x \in   (-Q/2, 0).                     $$ 
The basic interfaces are $x_0=0, x_1=-Q/2, x_2= -Q$.
  This time,   the amplitude equation is 
\begin{equation}\label{ZT008-4}
 -(\mu f_x)_x +  k^2 \mu f = \frac{1}{\sigma} U k^2 f \mu_x,  
 \quad \forall x \notin \{-Q,-Q/2, 0 \}.
\end{equation}

Inside the intermediar region, $\mu$  is a Heaviside  function. The 
derivative $\mu_x$  on the interface $x=-Q/2$ is a   Dirac distribution,
then
\begin{equation}\label{DIRAC}
 \int_{-Q}^0 \mu_x f^2 = f^2(x_1)(\mu_2-\mu_1). 
\end{equation}
The term \eqref{DIRAC} is not appearing  in   \cite{DAR-2}.   
We  multiply with $f$ the above relation  and  integrate on $(-Q,0)$,
then it follows
$$  - \int_{-Q}^{-Q/2} (\mu f_x f)_x  - \int_{-Q/2}^{0} (\mu f_x f)_x
+   \int_{-Q}^0 \mu f_x^2 +     k^2 \int_{-Q}^0 \mu f^2 =                 
   \frac{k^2U}{\sigma} f^2(x_1)(\mu_2-\mu_1).                            $$  
We use the notations $(FG)(x):=F(x)G(x)$, $ \quad f_i:=f(x_i)$ and get
$$
(\mu^+f_x^+ f)(-Q) - ( \mu^-f_x^-f )(-Q/2) +  
 (\mu^+f_x^+ f)(-Q/2) - ( \mu^-f_x^-f )(0)   +                            $$
$$+  I_1 + I_2 =     \frac{k^2U}{\sigma}  f^2(x_1)(\mu_2-\mu_1),           $$
\begin{equation}\label{PRE-SIGMA-AA}
  I_i= \mu_i \int _{x_i}^{x_{i-1}}(f_x^2+k^2f^2)dx, \quad   i=1,2.                 
\end{equation}
Recall 
$$ \mu^-(x_2)= \mu_W, \,\, \mu^+(x_0)= \mu_O, \quad 
   (f_x)^-(-Q)=kf_2, \,\, (f_x)^+(0)=-kf_0   .                             $$  
The  jump relations \eqref{LAPLACE002} in the points $a=x_2,x_1, x_0 $ are
$$
- {(\mu^+f_x^+)(x_2)} + {\mu_W k f_2}    
+ f_2\frac{Uk^2}{\sigma}[\mu_W -\mu_2] =  -  
\frac{T_2}{\sigma}f_2k^4,                                                   $$                                                                          
$$
-{(\mu^+f_x^+)(x_1)} + {(\mu^-f_x^-)(x_1)} =   
 f_1\frac{Uk^2 }{\sigma}[\mu_1-\mu_2]  -  
\frac{T_1}{\sigma}f_1k^4,                                                  $$                                                                          
\begin{equation}\label{PRE-SIGMA-AAB}
 {\mu_0kf_0} + {(\mu^-f_x^-)(x_0)}    
+ f_0\frac{Uk^2}{\sigma}[\mu_1 -\mu_O] =  -  
\frac{T_0}{\sigma}f_0 k^4.                                                       
\end{equation}                                                                         
In  \eqref{PRE-SIGMA-AA} - \eqref{PRE-SIGMA-AAB}   we use the  { \it viscosities
 $\nu_W, \nu_O$ } given in  \eqref{DIM-VISCO}, then $ \nu_i =b^2 \mu_i /12 $ and
it follows  
\begin{omitext}
$$
\sigma \{ {\mu_W k f^2_2}  
+ f_2^2 \frac{Uk^2}{\sigma}[\mu_W -\mu_2] +  \frac{T_2}{\sigma}f^2_2k^4 -
   f^2_1\frac{U}{\sigma}[\mu_1-\mu_2] + \frac{T_1}{\sigma}f^2_1k^2 +           $$     
\begin{equation}\label{PRE-SIGMA-AA2}
  {\mu_0kf^2_0}  
+ f^2_0\frac{Uk^2}{\sigma}[\mu_1 -\mu_O] +  \frac{T_0}{\sigma}f^2_0 k^4   +I_1+I_2    \}                                                                                                                             
=  k^2 U  f_1^2(\mu_2-\mu_1).                   
\end{equation}
\end{omitext}
$$
\sigma =  \frac{ S_0 f_0^2 + S_1 f_1^2 +  S_2 f_2^2 + k^2 U  f_1^2(\nu_2-\nu_1)}
{\nu_Ok f^2_0 + I_1+ I_2 +  \nu_W kf^2_1 },                                    $$
$$  S_i = Uk^2[\nu]_i  - k^4 {T_i} b^2/12, \quad   i=0,1,2,                   $$
$$ [\nu]_2= \nu_2-\nu_W, \quad  [\nu]_1= \nu_1-\nu_2, \quad
    [\nu]_0= \nu_0-\nu_1,                                                      $$
\begin{equation}\label{4-SIGMA}
  I_1 = \nu_1  \int _{x_1}^{x_0}(f_x^2+k^2f^2), \quad  
    I_2 = \nu_2  \int _{x_2}^{x_1}(f_x^2+k^2f^2).                              
\end{equation}

The following  dimensionless quantities  denoted by $'$ are introduced : 
$$ x'=x/Q, \, y' = y/Q,  \,  f'=f/U, \, \epsilon=b/Q \approx 10^{-3},        $$
$$ u^{'}=u/ U, \quad v'   =v/ U,  \quad T'= T /(\mu_W U)                     $$
\begin{equation}\label{coord-adim}
 \nu'= \nu/\mu_W ,  \quad   k'=kQ,  \quad  \sigma' = \sigma / (U/Q).    
\end{equation}

In the rest of this paper  we will omit the $'$. The dimensionless intermediate 
region is the interval $(-1,0)$.
The relation \eqref{4-SIGMA} gives us  the  dimensionless  growth rate, denoted 
by  $\sigma_2$ (recall $\mu_W=1$): 
$$  \sigma_2 =  
\frac{ \sum_{i=0}^{i=2}\{k^2[\nu]_i -\epsilon^2 k^4T_i / 12 \} f^{2}_i + 
  k^2  f_1^2[\nu]_1                                                          } 
{k\nu_O  f^{2}_0 + I_1+I_2 + k\cdot 1 \cdot f^{2}_2  }                       $$
$$ I_1= \nu_1 \int_{x_1}^{x_0} [ f_x^2 + k^2 f^2] dx, \quad 
I_2= \nu_2 \int_{x_2}^{x_1} [ f_x^2 + k^2 f^2] dx,                           $$
\begin{equation}\label{SIGMA-DOI-ADIM}
 x_2=-1, x_1=-1/2, x_0=0   .                     
\end{equation}
The factor  $\epsilon^2$  in front of the surface  tensions $ T_i $ is very important
for the stability analysis. In \cite{DAR-2} is given a similar formula, but with
dimensional quantities, then without the parameter $\epsilon$.

The dimensionless Saffman-Taylor growth rate and its maximal value are   obtained 
from \eqref{SIGMA_ST01}   and  \eqref{coord-adim}:
\begin{equation}\label{SIGMA_ST01_AD} 
\sigma_{D}=\frac{k (\nu_O - 1) - k^3 T \epsilon^2/12 }{\nu_O + 1} \leq
\sigma_{DM} = \frac{4(\nu_O-1)^{3/2}}{3(\nu_O+1) \epsilon \sqrt T}. 
\end{equation}

\begin{center}

{\bf  4.  The N-layer Hele-Shaw model  with  constant intermediate viscosities.} 

\end{center}

We divide the intermediate  region in $N$  small layers  with  equal length 
$(1/N)$. The interfaces are $x_0$ and $ x_i = -i/N, \quad  i=1,...N $. In 
the layer $(x_i, x_{i-1})$  we  consider the  constant viscosity  
$\nu(x)=\nu_i$ such that $\nu_0=\nu_O, \,\, \nu_{N+1}= 1$ (recall $\nu_W=1$) and 
\begin{equation}\label{STRATELE}
\nu_i = \nu_O - i(\nu_O-1)/(N+1), \quad (\nu^+ - \nu^-)_i=(\nu_0-1)/(N+1).                                       
\end{equation}          
The  amplitude   equations   are             
\begin{equation}\label{STRATELE-N}
 -(\nu f_x)_x +  k^2 \nu f = \frac{1}{\sigma} U k^2 f \nu_x,  
 \quad \forall x \notin \{-j/N \}, \quad  j=0,1,...N.
\end{equation}                               
The  growth  constants,  denoted  by   $\sigma_N$,  are  
 obtained just like  formula \eqref{SIGMA-DOI-ADIM}  in section 3:
\begin{equation}\label{SIGMA-N}
 \sigma_N = \frac{\sum_{i=0}^{i=N}\{  k^2(\nu^+ - \nu^-)_i -k^4 T_i\epsilon^2/12  \} f_i^2
+ \sum_{i=1}^{i=N-1} k^2(\nu^+ - \nu^-)_if_i^2   }
{  k\nu_O f_0^2 +  \sum_{i=1}^{i=N} I_i+ k  f_N^2 },               
\end{equation} 
$$              I_i = \int_{x_i}^{x_{(i-1)}}  \nu_i (f_x^2 + k^2 f^2 ), 
\quad f_i=f(x_i).                                                                   $$

 Our growth constants are real. An instability result is obtained if  {\it only one} 
growth constant is positive. It is much more difficult to prove  a stability result: all growth 
constants  must be negative. In this paper  we consider some  particular eigenfunctions $f$ and 
analyse the corresponding growth rates given by \eqref{SIGMA-N}. We prove that
even  if the number of intermediate layers is very large, the maximum value of the growth constants
is  not so small (in a bouned range of $k$) and is not depending on the surface tensions.

\vspace{0.25cm}

{\bf 4.1   An  upper  bund of $\sigma_N$.}
From \eqref{SIGMA-N}  we see that $\sigma_N$  is negative
beyond a finite value of $k$, then the ``dangereous''  
wave numbers $k$ are bounded. 

{\it Lemma 1.} If $k \in [0,1] $ and $N = 1/(c-a)$ is large enough s.t. 
$k^2(c-a)^2 \approx 0$, then we have
\begin{equation} \label{INEQ003-A}
f(x)= e^{kx} \quad  \forall x \in  (a,c)    \Rightarrow  
J(a,c):=  \int_{a}^{c} ( f_x^2+ k^2 f^2 ) \approx  
(k^2/N) [ f^2(a)+ f^2(c) ]. 
\end{equation}

{\it Proof.} As $f(x)=e^{kx}$  we get $f_x^2+ k^2 f^2=(f_xf)_x$ and   
\begin{equation} \label{VAL-EXACTA}
 J(a,c)= (f_xf)(c)-(f_xf)(a)= k[f^2(c)-f^2(a)].
\end{equation}
We use the trapezoidal rule  for   $ F \in C^2(a,c) $: 
$$ {\int^c_a F(x) dx = \frac{c-a}{2}[F(a) + F(c)] - R, \quad  
R = \frac{(c-a)^3}{12} F''(\chi)}, \quad \chi \in (a,c).                           $$
Consider $F(x)=f^2(x)= e^{2kx}$, then $F''(x)=4 k^2e^{2kx}$ and for bounded $k$
and small enough $(c-a)$ we get an arbitrary small $R$. As
$I=2 k^2  \int_{a}^{c} f^2 $, we use \eqref{VAL-EXACTA} and we have to prove 
\begin{equation}\label{INEQ003-B}
 k[f^2(c)-f^2(a) ] \approx  k^2(c-a) [ f^2(a)+ f^2(c) ].                              
\end{equation}
For this, we neglect $k^2(c-a)^2$ and  use the first order Taylor  expansion  of 
$f(x)=e^{kx}$:
$$ f(c) \approx f(a) + kf(a)(c-a), \quad f^2(c) \approx f^2(a) + 2 k f^2(a) (c-a),  $$
\begin{equation}\label{TR-2}
f^2(c)-f^2(a) \approx   2 k f^2(a) (c-a), \quad 
f^2(c)+ f^2(a) \approx 2f^2(a) +   2 k f^2(a) (c-a).                                   
\end{equation} 
The approximation   \eqref{INEQ003-B}   is equivalent with    
$$ k \cdot 2k (c-a) \approx    k^2 (c-a) [2 + 2 k  (c-a)]                             $$
which  holds because   $k^3(c-a)^2 \leq k^2(c-a)^2 \approx  0 $.   
\hfill  $\square$  

%%%%%%%%%%%%%%%%%%%%%%%%%%%%%%%%%%%%%%%

\vspace{0.25cm}

 We use {\it Lemma 1} for computing the integrals $I_i$, with
$(a,c)=(x_i, x_{i-1})$, $i=1,2,.., N$.

We consider  $ k \in [0,1] $   and $N=10^{4}$, then $k/N \leq  10^{-4}$ 
 and  the approximation   \eqref{INEQ003-B} holds. From   \eqref{SIGMA-N}  we get
$$
\sigma_N \approx  \frac{ G_0f_0^2  + \sum_{i=1}^{N-1} G_if_i^2 + G_Nf_N^2  }
{ k\nu_W f_N^2 +  k\nu_O f_0^2 +  
\sum_{i=1}^{i=N} (k^2/N)\nu_i(f_{i-1}^2+f_i^2)},                    $$
$$    G_i= 2k^2(\nu^+ - \nu^-)_i -k^4 T_i\epsilon^2/12,  
\quad i=1,2,..., N-1,                                                $$
\begin{equation}\label{SIGMA-N-B}
G_j=  k^2(\nu^+ - \nu^-)_j -k^4 T_j\epsilon^2/12, \quad    
j=0, N.
\end{equation}
An important new element appearing in this formula is based on  the Dirac 
distributions correspondning to the derivative $\nu_x$ on the interfaces.
As a consequence, 
the ``middle'' terms  $G_i, 1\leq i \leq N-1$,  are larger, compared with
$G_0, G_N$. 

We recall th well-known inequality
\begin{equation}\label{INEQ}
B_i, x_i >0 \Rightarrow 
\min \{ \frac{A_i}{B_i} \} \leq 
 \frac{\sum_{i=0}^{i=M}A_ix_i}{\sum_{i=0}^{i=M}B_ix_i} 
\leq \max \{ \frac{A_i}{B_i} \}.    
\end{equation}                                        
From   \eqref{SIGMA-N-B}  and \eqref{INEQ} it follows 
$$ \sigma_N \leq 
\frac{2k^2(\nu^+ - \nu^-)_{N-1} -k^4 T_{min}\epsilon^2/12}
{k^2(\nu_N+\nu_{N-1})/N}.                                             $$ 
   As a consequence, in the range $k \in [0,1]$, the upper bound of $\sigma_N$ is 
{\it a polynomial of order 2},  and not of order 3 as in \cite{DAR-2} and 
\cite{SAFF-TAY}. We have 
$$\nu_N+\nu_{N-1}=(2N-1+3\nu_O)/(N+1),                                 $$
then  from the last  estimate we get    
\begin{equation}\label{UPB-TR}
 \sigma_N < 2(\nu_O-1)\frac{N}{2N-1+3\nu_O}- 
k^2 T_{min}\epsilon^2 \frac{N(N+1)}{12(2N-1 + 3\nu_O)},
\end{equation}
\begin{equation}\label{UPB-TR-B}
\sigma_M \leq \sigma_{NM} = 2(\nu_O-1)\frac{N}{2N-1+3\nu_O}.            
\end{equation}

The estimate \eqref{UPB-TR} holds    for  $k \in [0,1] $ and  large 
enough values of $N$, but his maximum value \eqref{UPB-TR-B} {\it is not 
depending on the surface tension $T_{min}$.} Here is a strong contradiction
with the Saffman-Taylor result: it is very  natural to have  an
improvement of stability when the surface tensions are large enough.
From this point  of view, the present multi-layer   model is wrong.
Moreover, the maximum value  \eqref{UPB-TR-B} can not be  arbitrary small  
for large $N$.

For  $k \in (0,1), N=10^{4}, T =1/ \epsilon^2$,
the relations   \eqref{SIGMA_ST01_AD}  and \eqref{UPB-TR-B}  give us  
\begin{equation}\label{COMPAR-A}
\sigma_{NM} \approx (\nu_O-1) = 99; \quad  
\sigma_{DM} =  \frac{4 (\nu_O-1)^{3/2}}{3(\nu_O+1)} \approx 13.3.
\end{equation}
Then the  Saffman-Taylor formula gives us a more stable displacement
for a large enough surface tension. 
From this point  of view, the multi-layer  Hee-Shaw model with constant 
intermediate viscosities is useless. Future research is needed to see
the  cause of the strong contradiction with the Saffman-Taylor result.

\newpage

\begin{center}

 {\bf  5. Conclusions}

\end{center}

The interface between two Newtonian immiscible fluids   in a rectangular Hele-Shaw cell is 
unstable  when the  displacing fluid is less viscous. A  surface tension on the interface 
can improve  the stability - see the formula  \eqref{SIGMA_ST01} .

 An intermediate fluid  with  a variable  viscosity  between the 
displacing fluids can minimize  the  Saffman-Taylor instability when the 
surface tensions are different from zero
 - see  the papers  \cite{GILJE}, ~\cite{GOR-HOM-1},  ~\cite{GOR-HOM-2},  
~\cite{SHAH},  ~\cite{SLOBOD},    ~\cite{UZOIGWE}. 

 A continuous function  can be approximated by a step function.
For this reason, the multi-layer Hele-Shaw model, consisting  of $N$ intermediate 
fluids with constant  viscosities was studied  in 
~\cite{DAR-2},  \cite {D6},  ~\cite{DAR-DING},  \cite {D7}.  Upper bounds of the 
growth rates   were obtained in these papers.
If  all surface tensions   verify some conditions,  
an arbitrary small (positive) upper bound of the growth rates can be obtained, if $N$ 
is large enough.

In this paper we study the multi-layer Hele-Shaw displacements in rectangular  cells. 
The  three-layer case is considered in section 2. We get a formula  of the 
growth rates in the fourth-layer case with constant intermediate  viscosity - see  
section 3.
We  use this result for  $N$ intermediate constant-viscosity layers  and get a new
 upper bound for the growth constants in section 4, by using the dimensionless
quantities    \eqref{coord-adim}. We prove that even if the number of intermediate
layers is very large, the maximum value of the growth constants is not so small - 
see \eqref{UPB-TR}.

The most important result of our paper is the upper bound \eqref{UPB-TR-B} of the
growth constant (which holds only for bounded $k$ and large $N$). 
This result is based on three new elements: the terms due to the Dirac
distributions $\mu_x$ on interfaces, the dimensionless quantities 
and the new estimate of the growth rates given in section 4.

The maximum   value of \eqref{UPB-TR-B} is not depending on the surface tensions and can 
not be arbitrary small  for a large   enough number of intermediate layers. 
Then we have a significant instability, even if the surface tensions on the interfaces
are very large. This contradiction with the Saffman-Taylor formula  
raises questions about the validity of the multi-layer model.

\end{document}